\documentclass[12pt]{article}

\usepackage{a4,graphicx,amsmath,amssymb,cite}
\usepackage[verbose]{wrapfig}
\def\Bbb{\mathbb}
\def\BZ{\Bbb Z} 
\def\BC{\Bbb C} 

\catcode`@=11 \@addtoreset{equation}{section} \catcode`@=12

\begin{document}
\begin{titlepage}
\renewcommand{\thefootnote}{\fnsymbol{footnote}}
\begin{flushright}
IITM/PH/TH/2007/1  \\
October 2007
\end{flushright}
\vspace{1.0cm}
\begin{center}
\large{\bf A note on perturbative aspects of Leigh-Strassler
deformed $\mathcal{N}=4$ SYM theory}
\end{center} 
\bigskip 
\begin{center}
Kallingalthodi Madhu\footnote{E-mail: \texttt{madhu@physics.iitm.ac.in}} and 
Suresh Govindarajan\footnote{E-mail: \texttt{suresh@physics.iitm.ac.in}} 
\\  Department of Physics \\
Indian Institute of Technology Madras 
\\ Chennai 600036 INDIA\\ 
\end{center}

\begin{abstract}
We carry out a perturbative study of the Leigh-Strassler deformed 
$\mathcal{N}=4$ SYM theory in order to verify that the trihedral 
$\Delta(27)$ symmetry holds in the quantum theory. We show that 
the $\Delta(27)$ symmetry is preserved to two loops (at finite 
$N$) by explicitly 
computing the superpotential. The perturbative superpotential is 
not holomorphic in the couplings due to finite contributions. 
However, there exist coupling constant redefinitions that restore 
holomorphy. Interestingly, the same redefinitions appear (in the 
work of Jack, Jones and North) if one requires the three-loop 
anomalous dimension to vanish in a theory where the one-loop 
anomalous dimension vanishes. However, the two field 
redefinitions seem to differ by a factor of two.
\end{abstract}
\end{titlepage}
\renewcommand{\thefootnote}{\arabic{footnote}}
\setcounter{footnote}{0}

\section{Introduction}

The Leigh-Strassler deformations of the $\mathcal{N}=4$ SYM 
theory\cite{Leigh:1995ep} are a class of $\mathcal{N}=1$ super 
conformal field theories that are particularly interesting in the 
context of AdS/CFT correspondence. Though the gravity dual for 
the most general deformation of $\mathcal{N}=4$ theory is not yet 
known, a subclass of deformations known as $\beta$-deformation 
has been well studied 
\cite{Lunin:2005jy,Freedman:2005cg,Penati:2005hp,Mauri:2005pa}. 
There have been perturbative studies of the states of the super 
conformal algebra, especially the chiral primary states 
\cite{Mauri:2006uw,Madhu:2007ew}. The Leigh-Strassler (LS) theory 
is conformal on a subspace of the coupling space defined by the 
matter couplings $Y_{IJK}$ and the gauge coupling $g$. The 
conformal properties can be understood by studying the 
$\beta$-functions for the chiral couplings and the NSVZ 
$\beta$-function for the gauge coupling. The superpotential of 
the theory is protected from renormalizations by holomorphy. 
Hence, for the chiral couplings the $\beta$-functions must be 
proportional to the anomalous dimensions of the chiral 
superfields.
\begin{equation}
\beta_{IJK}\equiv 
\beta(Y_{IJK}) \sim Y_{LJK}\ \gamma^{L}_I + Y_{ILK}\ \gamma^{L}_J
+Y_{IJL}\ \gamma^{L}_K\ , 
\label{betayukawa}
\end{equation}
For LS theory, the gauge $\beta$-function given by NSVZ 
\cite{Novikov:1985rd} reduces to
\begin{equation}
\beta^{NSVZ}(g)=-\frac{g^3}{32\pi^2}\Big[\frac{2N\gamma^I_I}{(N^2-1)(1-g^2N(16\pi^2)^{-1}}\Big]\ .
\end{equation}
This is again proportional to the anomalous dimensions. From Eqn. 
\eqref{betayukawa}, the anomalous dimension matrix seem to have 
nine components for a theory with three flavors like the LS 
theory. However as we shall see, the symmetry of the LS action 
constrains this matrix to be proportional to unit matrix, giving 
rise to a single condition $\gamma = 0$ which defines the 
subspace on which the theory is conformal.

Classically, the LS theory has a discrete non-abelian symmetry 
given by trihedral $\Delta(27)$ group 
\cite{Aharony:2002hx,Wijnholt:2005mp}. In ref. 
\cite{Madhu:2007ew} it was shown that the chiral primaries can be 
classified as representations of this $\Delta(27)$ group. Thus it 
is important to know whether the $\Delta(27)$ group is a symmetry 
of the quantum theory. For this, one has to show that the quantum 
corrected superpotential and the K\"ahler potential preserves 
this symmetry. Another aspect which is quite interesting to 
understand is whether conformal invariance and holomorphy of the 
theory is preserved quantum mechanically. In the computation of 
anomalous dimension of scalar composite operators we find that 
the contribution from the non $F$-terms cancel when we impose the 
condition for conformal invariance. This suggests that conformal 
invariance of the LS theory may be sufficient to ensure 
holomorphicity.

In the following section we explain the Leigh-Strassler theory 
and its symmetries and write the superpotential in a useful form. 
We discuss in section 3, the role of $\Delta(27)$ in preserving 
the conformal invariance of the theory by studying the anomalous 
dimension. In section 4, we check the conformal properties of the 
theory by computing anomalous dimension upto three-loop, 
following ref. \cite{JJNa} and point out the existence of 
coupling constant redefinitions that preserve conformal 
invariance of the theory. Section 5, explains computation of 
two-loop effective superpotential. The two-loop contribution is 
not holomorphic in coupling constant $h$, as is expected with a 
1PI effective superpotential. We point out that it is strikingly 
similar to the three-loop contribution to the anomalous dimension 
giving rise to the possibility that same field redefinitions 
preserve conformal invariance and holomorphy of the LS theory. We 
also briefly describe the two-loop effective K\"ahler potential 
in section 6 and show that it preserves the $\Delta(27)$ symmetry 
of the LS theory. We conclude in section 7 with remarks about the
results of the paper. We give the details of our computations in the 
various appendices.

\section{LS deformed ${\cal N} = 4$  Yang-Mills theory}

The Lagrangian density of the Leigh-Strassler theory in terms of
${\cal N}=1$ superfields is
\begin{eqnarray}
\label{zansup}
&{\cal L}& =\ \int d^2 \theta d^2 \bar{\theta}\ \textrm{Tr}\Big( e^{-g V} 
\bar{\Phi}_i e^{g V} \Phi_i \Big) + \Big\{ \frac{1}{2 g^2} \int d^2 \theta 
\ \Big[\textrm{Tr} \Big( \mathcal{W}^{\alpha} \mathcal{W}_{\alpha} \Big)  \\
&+& i h \textrm{Tr} \Big( e^{i \pi \beta} \Phi_1 \Phi_2 
\Phi_3
 - e^{- i \pi \beta} \Phi_1 \Phi_3 \Phi_2 \Big) 
+ \frac{ih'}{3}  \textrm{Tr} \Big( \Phi_1^3 + \Phi_2^3
+ \Phi_3^3 \Big) \Big] + h.c.\Big\} \nonumber 
\end{eqnarray}
All fields transform in the adjoint of $SU(N)$ and we assume that 
$N>2$. Let $q\equiv e^{i\pi\beta}$ and $\bar{q}\equiv e^{-i\pi 
\beta}$. When $\beta$ is real, then $q$ and $\bar{q}$ are complex 
conjugates of each other. The imaginary part of $\beta$ can 
always be absorbed by a redefinition of $h$. We have also set 
$\Theta=0$.

\noindent The theory has the symmetry of the trihedral group, 
$\Delta(27)\sim ((\BZ_3)_R\times \BZ_3)\rtimes{\cal C}_3$, which 
is a discrete non-Abelian subgroup of $SU(3)\subset SU(4)$ 
\cite{Aharony:2002hx}. This is obtained from the $\beta$-deformed 
theory by further breaking down the $U(1)^3$ symmetry. The action 
of $\Delta(27)$ on the fields of the theory is as follows:
\begin{eqnarray}
h &:& \Phi_1 \longrightarrow \Phi_1 \ , \Phi_2 \longrightarrow \omega\ \Phi_2 \ ,
\Phi_3 \longrightarrow \omega^2\ \Phi_3 \nonumber \\
\tau &:& \Phi_1 \longrightarrow \Phi_2  \longrightarrow \Phi_3
\longrightarrow \Phi_1 \nonumber
\end{eqnarray}
where $h$ generates $\BZ_3$ and $\tau$ generates ${\cal C}_3$ and 
$\omega$ is a non-trivial cube-root of unity. Further $(\BZ_3)_R$ 
is a sub-group of $U(1)_R$ -- we assign charge $+1$ to all fields 
(this is $3/2$ times their $R$-charge).

We can rewrite the superpotential by combining the three chiral 
superfields into one superfield and use one meta-index 
$I,J,K,L\ldots$ representing the $SU(N)$ adjoint index 
$a,b,c,d,\ldots$ as well as the index $i,j,k,l,\ldots=1,2,3$ 
which labels the three chiral superfields. The Leigh-Strassler 
superpotential (the trace below is in the fundamental 
representation of $SU(N)$)
\begin{equation}\label{cubicsup}
W_{LS} = \frac{f}{6} \epsilon^{ijk}\textrm{Tr}_F (\Phi_i\Phi_j\Phi_k)
+\frac{1}{6} c^{ijk}\textrm{Tr}_F (\Phi_i\Phi_j\Phi_k)\ ,
\end{equation}
where the fully symmetric tensor $c^{ijk}$ is given by
\begin{equation}
c^{ijk}=\left\{ \begin{array}{ll} c_0,\ & i\neq j\neq k \neq i, \\
                                  c_1,\ & i=j=k, \\
                                  0,\   & \textrm{otherwise.}
 
                \end{array}\right. \ .
\label{cijk}
\end{equation}
One can prove that \textit{only} the above choice for $c^{ijk}$ 
leads to a superpotential that is invariant under the trihedral 
group $\Delta(27)$. In particular, couplings such as $c^{112}$ 
vanish and $c_1=c^{111}=c^{222}=c^{333}$. Thus, if $\Delta(27)$ 
is to remain of symmetry of the quantum theory, such couplings 
must \textit{not} arise in the quantum 
theory\cite{Aharony:2002tp}.

In order to be able to compare with the usual representation of 
the LS superpotential, we give the relationship to the usual 
parameters $h$, $q$, $h'$:
\begin{equation}
f=h(q+\bar{q})\quad,\quad c_0=h(q-\bar{q})\quad,\quad c_1=2h'\ .
\end{equation}
In terms of the meta-index, the LS superpotential can be written as
follows(matching the notation of \cite{JJNa}):
\begin{equation}
W_{LS}=\frac16 Y^{IJK}\Phi_I\Phi_J\Phi_K\ ,
\end{equation}
where
$$
Y^{IJK} \equiv Y^{(ia)(jb)(kc)} = \frac1{2}\Big(i f 
\epsilon^{ijk}\otimes f_{abc} + 2 c^{ijk}\otimes d_{abc}\Big)\ .
$$
The generators of $SU(N)$ in the fundamental representation have been
taken to satisfy the identity (with the normalization
$\textrm{Tr}_F(T_aT_b)=\delta_{ab}$)
\begin{equation}
\textrm{Tr}_F\big(T_aT_bT_c\big)\equiv \frac12 \big[i f_{abc} + 2 
d_{abc}\big]\ .
\end{equation}
$f_{abc}$ are the structure constants of $SU(N)$ and $d_{abc}$ is the totally
symmetric tensor.

It is interesting to observe the quantum mechanical properties of 
the measure in the LS theory before we begin our discussion of 
perturbative properties. As shown in 
\cite{Arkani-Hamed:1997mj,Arkani-Hamed:1997ut}, the NSVZ 
$\beta$-function \cite{Novikov:1985rd} can be viewed as arising 
from the non-trivial transformation of the measure of the path 
integral under rescaling of the chiral and vector superfields. 
For theories with matter fields in three flavors in the adjoint 
adjoint representation, the $\beta_{NSVZ}$ is proportional to the 
anomalous dimension $\gamma$ of the chiral superfield. The 
requirement of vanishing of the $\gamma$-function defines the 
subspace of the space of couplings where the theory remains 
conformal. Particularly interesting is the question of how the 
measure changes under the $\Delta(27)$ action. The measure of 
$\mathcal{N}=4$ SYM theory is invariant under $SU(4)_R$. As the 
spectrum of the LS theory is identical to that of $\mathcal{N}=4$ 
SYM theory, it must also be invariant under the action of 
trihedral group $\Delta(27)$ which is after all a subgroup of 
$SU(4)_R$.

\section{Conformal invariance of the LS theory}

The trihedral symmetry group, $\Delta(27)$, can be seen as a 
finite sub-group of $SU(3)\subset SL(3,\BC)$. An arbitrary 
gauge-invariant cubic superpotential involving three chiral 
superfields (transforming in the adjoint of $SU(N)$), $\Phi^i$, 
consists of eleven independent (complex) couplings. Linear 
redefinitions of the three fields form the group $SL(3,\BC)$ 
while $SU(3)$ is the sub-group of $SL(3,\BC)$ which preserves the 
(diagonal) kinetic energy which is encoded in the tree-level 
K\"ahler potential $\bar{\Phi}^i\Phi_i$. By means of linear 
redefinitions, it is possible to set eight of the eleven 
couplings that appear in the superpotential to zero and obtain 
the form given in Eqn. (\ref{cubicsup}). The trihedral group 
$\Delta(27)$ emerges as the subgroup of $SL(3,\BC)$ that 
preserves that form. If the K\"ahler potential also retains its 
diagonal form, then $\Delta(27)$ is a symmetry of the theory.

We will now show that the trihedral symmetry and gauge-invariance 
implies that $\gamma^I_J\propto \delta^I_J$. Recall that the only 
gauge-invariant $SU(N)$ tensor is $\delta_a^b$. Thus the 
gauge-invariance requires that the matrix of anomalous dimensions 
be proportional to $\delta^a_b$. Thus, we write
$$
\gamma^{ia}_{jb}\equiv \gamma^i_j
\delta^a_b\ ,
$$ 
where we have separated the flavor indices from the gauge indices.

Recall, that invariance under $\Delta(27)$ implies that couplings 
such as $c^{112}$ vanish and requires $c^{111}=c^{222}=c^{333}$.  
For this to remain so we need $\beta(c^{112})=0$ and 
$\beta(c^{111})=\beta(c^{222})$ to all orders in the quantum 
theory. Consider $\beta(c^{112})$ -- it is given by (using 
$Y^{1a~1b~2c}\sim c^{112}d_{abc}$)
\begin{equation}
\beta\big(Y^{1a~1b~2c}\big)\sim d_{abc}\Big(c^{11k}\gamma_{k}^{2} + 2 
c^{1k2}\gamma_{k}^{1}  \Big)\ .
\end{equation}
The vanishing of the RHS in the background values of $c^{ijk}$ 
given in Eqn. (\ref{cijk}) needs $\gamma^2_1=0$ and 
$\gamma^1_3=0$. Similarly, one can show that all off-diagonal 
terms vanish by considering the $\beta$ functions for all 
$c^{iik}$ with $i\neq k$. We still need to show that the diagonal 
matrix is proportional to the identity matrix. For this we 
consider
\begin{equation}
\beta\big(Y^{1a~1b~1c}\big)-\beta\big(Y^{2a~2b~2c}\big) 
\sim d_{abc} \big(\gamma_1^1c^{111}-\gamma_2^2c^{222}\big)\ .
\end{equation}
This vanishes only when $\gamma_1^1=\gamma^2_2$. Similar 
considerations also require $\gamma_1^1=\gamma_3^3$. This 
completes the proof that $\gamma^i_j\propto \delta^i_j$. We can 
thus write
\begin{equation}
\label{anommatrixprop}
\gamma^I_J \equiv  \gamma\ \delta_I^J\ .
\end{equation}
Thus, the vanishing of all the
$\beta$-functions imposes only \textit{one} condition, i.e., 
$$
\gamma(g,h,\beta,h')=0\ ,
$$
in the space of coupling constants in the LS theory. Below, we 
explicitly verify that the matrix of anomalous dimensions 
satisfies Eqn. (\ref{anommatrixprop}) to three loops by 
specializing the results of Jack, Jones and North(JJN) to the LS 
theory\cite{JJNa}.

\section{Computing the anomalous dimension}

We write the $\gamma$ function (anomalous dimension) as
\begin{equation}
\gamma = \gamma^{(1)} + \gamma^{(2)} + \gamma^{(3)} + \cdots
\end{equation}
where the superscript denotes order of the loop contribution. The
answers are given in the $\overline{MS}$-scheme. 

One has the following general expressions for $\gamma^{(1)}$ and 
$\gamma^{(2)}$\cite{Parkes:1984dh,Parkes:1985hj,West:1984dg,Jones:1983vk,Jones:1984cx}: 
We follow the notation of JJN except that our gauge coupling 
constant $g$ is $\sqrt2$ times theirs\cite{JJNa}.
\begin{eqnarray}
(16\pi^2) {\gamma^{(1)}}^I_J &=& \frac12 Y^{IKL}Y_{JKL}-g^2 C(R)^I_J \equiv 
P^I_J\\
(16\pi^2)^2 {\gamma^{(2)}}^I_J &=& \Big(Y^{IMK}Y_{JMN}-g^2 C(R)^K_J \delta^I_N
\Big) P^N_K+g^4 C(R)^I_J Q
\end{eqnarray}
where $Y_{IJK}=(Y^{IJK})^*$, $Q=T(R)-3C(G)$. We define $C(G) 
\delta^a_b = f^{acd} f_{bcd} = 2 N \delta^a_b$, $T(R) \delta_{ab} 
= \mathrm(T_a T_b) $ and $C(R)^I_J = (T_a T_a)^I_J$. Here $R$ 
refers to the reducible representation given by three copies of 
the adjoint representation. Specializing the the LS theory where 
$Q=0$ and
\begin{align}
\frac12Y^{IKL}Y_{JKL} &=  \frac{1}{2} N\delta^{I}_J 
\Big[|f|^2  +  \big( |c_0|^2 + \tfrac{|c_1|^2}{2}\big)\tfrac{N^2-4}{N^2}\Big] 
\ ,\\
&=  2 N\delta^{I}_J\Big[ |h|^2 
 - |h|^2 \frac{|q-\bar{q}|^2}{N}+|h'|^2\tfrac{N^2-4}{2N^2} \Big]
\equiv \hat{P}\ \delta^I_J 
\end{align}
The one-loop $\gamma$ function for the fields is given by JJN to be
(using $C(R)^I_J = 2N\delta^I_J$)
\begin{equation}
16\pi^2 {\gamma^{(1)}}^I_J =16\pi^2 {\gamma^{(1)}}\ \delta^I_J =(\hat{P} - 2 g^2 N)\delta^I_J\ .
\end{equation}
The vanishing of the one-loop $\gamma$ function is then
\begin{align}
\label{oneloopgamma}
\gamma^{(1)}=0 \implies  
\boxed{
N \Big[ |h|^2 
 - |h|^2 \frac{|q-\bar{q}|^2}{N}+|h'|^2\tfrac{N^2-4}{2N^2} \Big] -  g^2 N=0\ . 
}
\end{align}
In the $\mathcal{N}=4$ limit, this expression simplifies to 
$g^2=|h|^2$ and also matches the expression given by Penati et. 
al.\cite{Penati:2005hp}. The two-loop correction is given by
\begin{equation}
(16\pi^2)^2\ {\gamma^{(2)}}^I_J =
\big[-2\hat{P}-2g^2N\big]\big[\hat{P}-2g^2N\big]\delta^I_J \ ,
\end{equation}
also vanishes in the sub-space where $\gamma^{(1)}=0$.  This is 
the well-known result that one-loop finite theories are two-loop 
finite as well.

\begin{figure}[ht] 
\centering 
\includegraphics[height=1.3in]{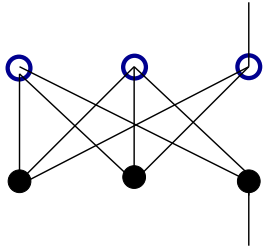} 
\caption{Non-planar contribution to $\gamma^{(3)}$ -- a filled 
circle represents the chiral cubic vertex and a open circle 
represents an anti-chiral vertex.}

\label{nonplanar}
\end{figure}

The three-loop $\gamma$-function does \textit{not} vanish in the 
$\overline{MS}$ scheme. It was computed by JJN who also showed 
that there exists a renormalization scheme wherein the three-loop 
gamma function vanishes provided the one-loop contribution does. 
In the $\gamma^{(1)}=0$ sub-space, Parkes computed the three-loop 
gamma function\cite{Parkes:1985hh}

\begin{align} (16\pi^2)^3 
{\gamma_P^{(3)}}_I^J&= \kappa \frac{g^6}{2^3} \big[12 C(R)C(G)^2 
- 2 C(R)^2 C(G)-10C(R)^3-4C(R)\Delta(R)\big] \nonumber \\ 
&+ \kappa \frac{g^4}{2^2} \big[4C(R)S_1-C(G)S_1 + S_2-5S_3 \big] - 
\kappa \frac{g^2}2 Y^*S_1Y+\kappa\frac{M_I^J}4 
\end{align} 
where $\kappa=6\zeta(3)$ and 
\begin{eqnarray*} 
{S_1}_I^J&=&Y_{IMN}\ 
C(R)^M_P\ Y^{JPN} = 4 N \hat{P}\delta_I^J \ ,\\ 
(Y^*S_1Y)_I^J&=& 
Y_{IMN}\ {S_1}_P^M\ Y^{JPN} = 8N^2 \hat{P}^2 \delta_I^J\ , \\ 
{S_2}_I^J&=&Y_{IMN}\ C(R)_P^M C(R)_Q^N\ Y^{JPQ}=8N^2 \hat{P} \delta_I^J\ ,\\ 
{S_3}_I^J&=& Y_{IMN}\ (C(R)^2)_P^M\ Y^{JPN}=8 N^2 \hat{P} \delta_I^J\ ,\\ 
\Delta(R) &=& \sum_\alpha C(R_\alpha)T(R_\alpha)=12N^2\ ,\\ 
M_I^J &=& Y_{K_1K_2K_3}Y_{L_1L_2L_3}Y_{IM_1M_2} Y^{JK_3L_3}Y^{K_1L_1M_1} 
Y^{K_2L_2M_2}\ . 
\end{eqnarray*} 
Above, we have given the values taken by the various terms for 
the LS theory except for $M_I^J$ which involves a complicated 
expression and is given later. Putting in these expressions, we 
find that all $g$-dependent terms vanish in the $\gamma^{(1)}=0$ 
subspace leaving behind a simple expression:
\begin{eqnarray} (16\pi^2)^3 
{\gamma_P^{(3)}}_I^J  = \frac{\kappa}4 M_I^J\ , 
\end{eqnarray} 
This is indeed an interesting result -- it implies that (in the 
$\gamma^{(1)}=0$ subspace) the only diagram which contributes to 
$\gamma^{(3)}$ in the LS theory is the only \textit{non-planar} 
diagram (see Figure \ref{nonplanar}) that first appears at 
three-loop. This diagram vanishes in $\mathcal{N}=4$ SYM theory.

An explicit computation reveals that $M_I^J$ is indeed proportional to the
identity matrix(see Appendix B  for more details)
\begin{align*}
M_I^J=&\frac{3\zeta(3)}2\frac{4-N^2}{N(N^2-1)}\Big[
\frac12\Big(18|c_0|^2|c_1|^2+2c_0^3(2\bar{c}_0^3
+\bar{c}_1^3)+c_1^3(2\bar{c}_0^3
+\bar{c}_1^3)\Big)\left(1-\tfrac{10}{N^2}\right)
\\
&
+ \Big(4 \bar{f}^2(4c_0^3 \bar{c}_0+2c_1^3\bar{c}_0-6c_0^2|c_1|^2)
+4 f^2(4\bar{c}_0^3 c_0+2\bar{c}_1^3c_0-6\bar{c}_0^2|c_1|^2)
\Big)
\Big]\delta_I^J\ .
\end{align*}
The above term clearly vanishes in the $\mathcal{N}=4$ limit and also
vanishes in the large-$N$ limit reflecting the non-planar nature of
the diagram. 

\subsection{Coupling constant redefinitions}

In ref. \cite{JJNa}, Jack, Jones and North have an interesting 
observation. They show that there exists a redefinition of the 
coupling constants for which the three-loop $\gamma$ function 
also vanishes in a theory where $\gamma^{(1)}=0$. This is 
equivalent to moving away from the $\overline{MS}$ scheme. For 
the LS theory, due to the additional cancellations that we 
observed, the redefinition is simpler than the one used by JJN. 
One needs
\begin{equation}
\label{redefone}
(16\pi^2)^2\ \delta Y_{IJK} =\frac{\kappa}4\mathcal{M}_{IJK} 
\end{equation}
where
\begin{equation}
\label{MIJK}
\mathcal{M}_{KLM}=Y^{I_1I_2I_3}Y^{J_1J_2J_3}Y_{I_1J_1K}Y_{I_2J_2L}Y_{I_3J_3L}
\end{equation}
On carrying out the coupling constant redefinition, the condition for
conformal invariance continues to be the one given in Eqn. 
(\ref{oneloopgamma}) albeit in the redefined couplings.

\section{Two-loop effective superpotential}

We next move on to the computation of the effective 
superpotential to two-loops. It was shown by West that in 
theories with massless fields such as the cubic Wess-Zumino 
model, that the 1PI superpotential is non-holomorphic in coupling 
constants due to finite contributions\cite{Jack:1990pd}. Such 
contributions do arise in our theory as well. We work out the 
coupling constant redefinition that is required to restore 
holomorphy. It turns out to be identical in structure to the one 
given in Eqn. (\ref{redefone}) but is twice as large.

Below we give all the diagrams which can potentially contribute 
to the superpotential at two-loops. Diagrams (a)-(d) contribute 
terms that are proportional to the tree-level superpotential 
while (e) vanishes. All these diagrams also contribute to the 
$\mathcal{N}=4$ theory. Diagram (f) is non-planar and leads to a 
non-holomorphic contribution to the superpotential.
\begin{figure}[ht]
\centering
\includegraphics[height=3.0in]{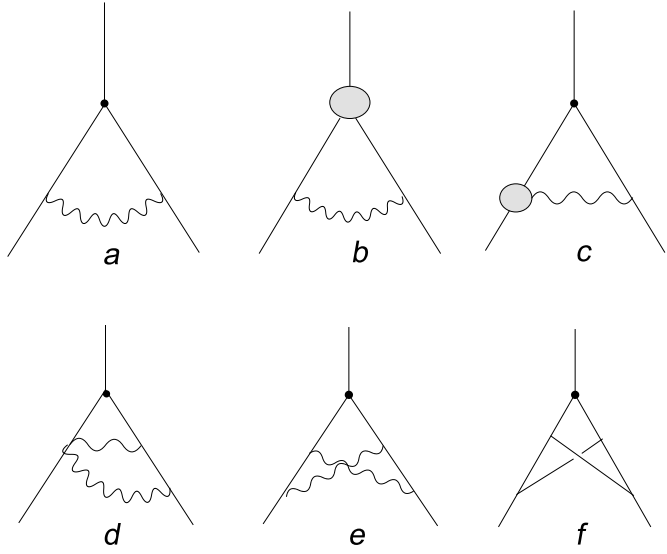}
\caption{Contributions to the two-loop effective action. The blob that
appear in (b) and (c) are one-loop vertex corrections.
}\label{EAD}
\end{figure}
All the diagrams above lead to finite integrals. For details of 
evaluation of these diagrams we refer to appendix C.

\subsection{The non-planar diagram}

The effective superpotential thus obtains a non-trivial
contribution only from the diagram 
\begin{figure}[ht]
\centering
\includegraphics[height=2.4in]{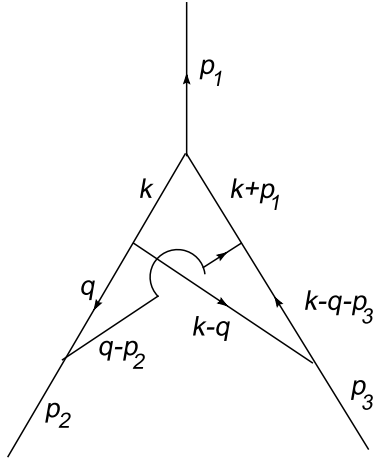}
\caption{Chiral contribution to superpotential}\label{EAF}
\end{figure}
\begin{align}
&\frac{1}{6^5} \times \frac{(3!)^5}{3! 2!} \times 
\mathcal{M}^{IJK}
\int d^2 \theta_1 d^2 \theta_2  d^2 \theta_3 d^2 \bar{\theta}_4 
d^2 \bar{\theta}_5 \nonumber \\
& \int\frac{d^Dk d^Dq}{(2 \pi)^{2D}} 
\frac{\Phi_I(p_2 + p_3, \theta_1) \Phi_J(-p_2 , \theta_2) \Phi_K(-p_3, \theta_3)}{k^2 q^2 (k-q)^2 (q-p_2)^2
(k-q-p_2)^2 (k-p_2-p_3)^2} \nonumber \\
& \bar{D}_1^2 D_4^2 [k] \delta^4(\theta_{14}) \;
\bar{D}_2^2 D_4^2 [q] \delta^4(\theta_{24}) \; \bar{D}_3^2 D_4^2 [k-q] \delta^4(\theta_{34}) \; \\
&\bar{D}_1^2 D_5^2 [k-p_2-p_3] \delta^4(\theta_{15}) 
 \bar{D}_2^2 D_5^2 [q-p_2] \delta^4(\theta_{25}) \; \bar{D}_3^2 D_5^2 [k-q-p_3] \delta^4(\theta_{35}) \nonumber \ ,
\end{align}  
where $\mathcal{M}^{IJK}$ has been defined in Eqn. (\ref{MIJK}). 
Note that the momentum in the square brackets in the last two 
lines indicate the momentum appearing in the superderivatives. 
Details like the algebra of $D$-operators and simplification of 
the flavor and color factors in this computation are provided in 
the appendices B and C. We obtain the two-loop correction to the 
superpotential as
\begin{eqnarray}
\mathcal{K}\
\frac{M^{IJK}}{12}  \int d^2\theta   \; \Phi_I(p_3, \theta) \Phi_J(p_3 , 
\theta) \Phi_K(-p_3, \theta)
\end{eqnarray}
where $\mathcal{K}$ is the finite integral
\begin{equation}
\mathcal{K}\equiv 
p_3^2\int \frac{d^Dk d^Dq}{(2 \pi)^{2D}}
\frac1{k^2 r^2  (k-r)^2 (r-p_3)^2 (k-p_3)^2} 
= \frac{\kappa}{(16 \pi^2)^2}\ 
\end{equation}
with $\kappa=6\zeta(3)$.
Putting in the explicit form of $\cal{M}^{IJK}$ for the LS superpotential we
obtain
\begin{align}
\delta c_{1}  
&=\mathcal{K} \Big[\tfrac{-N^2+10}{2 N^2}\Big] \Big[6|c_0|^4 c_1+\bar{c}_1^2(2c_0^3+c_1^3)
\Big]
- \mathcal{K} \Big[6 \bar{f}^2 c_0^2 c_1\Big]
+ 3\mathcal{K}\Big[2 f^2 (\bar{c}_1^2c_0-\bar{c}_0^2c_1)\Big] \nonumber\\
\delta c_{0}&=
\mathcal{K} \Big[\tfrac{-N^2+10}{2N^2}\Big] \Big[\bar{c}_0(6|c_1|^2c_0^2 +
\bar{c}_0 (2c_0^3+c_1^3)) \Big]
+ \mathcal{K} \bar{f}^2 (2c_0^3 +c_1^3)
+ 6\mathcal{K}\Big[ f^2 \bar{c}_0(|c_0|^2-|c_1|^2)\Big] \nonumber\\
\delta f &= \mathcal{K} \Big[\tfrac{-N^2+4}{2N^2}\Big] 
\Big[\bar{f}(-3 |c_1|^2c_0^2+\bar{c}_0(2c_0^3+c_1^3))\Big]
\end{align}
Specialising the above result to the $\beta$-deformed theory, it 
simplifies to the one given in the two-loop computation in ref. 
\cite{Mauri:2006uw}(except for a mismatch of a factor of two).

\subsubsection{Coupling constant redefinitions in the two-loop superpotential}

The two-loop contribution to the effective superpotential thus leads to
a redefinition of the form
\begin{equation}
\label{redeftwo}
(16\pi^2)^2\ \delta Y^{IJK} =\frac{\kappa}2 \ \mathcal{M}^{IJK}\ .
\end{equation}
Holomorphy in the couplings is restored if we make a redefinition 
of the $Y^{IJK}$ to absorb the non-holomorphic pieces in 
$\mathcal{M}^{IJK}$. We can compare this redefinition with the 
one required to make the gamma function vanish to three-loops 
given in Eqn. (\ref{redefone}). It is interesting to note that 
both are proportional to $\kappa \mathcal{M}^{IJK}$ but 
\textit{differ} by a factor of two. The result of 
\cite{Mauri:2006uw} however requires the same redefinition -- we 
have however been unable to find an error, if any, in our 
computation.

One may wish to know whether it is truly essential for the two 
redefinitions to agree. In principle, there is no such 
requirement. We could insist on holomorphy in couplings and 
choose the redefinition that is required by it. As the 
redefinition for conformal invariance is different, it implies 
that the condition of conformal invariance obtains a correction 
at three-loop and finite-$N$. So if one wishes to preserve the 
one-loop conformal invariance condition, then one needs to give 
up holomorphy in the couplings.

\section{One-loop effective K\"ahler potential}

The K\"ahler potential for any $\mathcal{N}=1$ supersymmetric theory is
non-holomorphic and provides the kinetic terms as well as the
interactions between vector superfields with the chiral superfields.
At tree-level in the LS theory, we have chosen the K\"ahler potential
$\Phi_i\bar{\Phi}^i$. The effective one-loop K\"ahler
potential has been computed in \cite{Pickering:1996gt,Grisaru:1996ve}
and we make use of their results -- our notation is adapted from the
second reference. In the Feynman gauge, the one-loop K\"ahler potential
is given by
\begin{equation}
\label{Keff}
K_{\textrm{eff}}^{\textrm{1-loop}} = 
\sum_{n=1}^{\infty}\int \frac{d^4kd^4\theta
}{(2\pi)^4}\ \frac{(-1)^{n+1}}{2n\ k^{2n+2}} \
\textrm{Tr}\Big(\big[\bar{\mu}\mu\big]^n
-2M^n\Big)
\end{equation}
where the first contribution arises from the insertion of $n$ chiral and
anti-chiral vertices and the second contribution arises from the
insertion of $n$ interaction vertices involving the gauge field and
and the scalars. We have defined
\begin{equation}
\mu^{IJ}= Y^{IJK} \mathbf{\Phi}_K \quad,\quad
\bar{\mu}_{IJ}= Y_{IJK} \mathbf{\bar{\Phi}}^K \quad,\quad
M_{ab}= \frac{g^2}2 \mathbf{\bar{\Phi}}^l\{T_a,T_b\}\mathbf{\Phi}_l\ ,
\end{equation}
with the boldface $\mathbf{\Phi}$ indicating that the computation is
being carried out in the background given by  $\mathbf{\Phi}$.

The first term in Eqn. (\ref{Keff}) is logarithmically divergent in the
UV and is proportional to
\begin{equation}
\frac12 \textrm{Tr}(\bar{\mu}\mu) - \textrm{Tr}(M) =
%(\frac12 Y_{IJK-g^2 C(G))\ \mathbf{\bar{\Phi}}^L \mathbf{\Phi}_L =
(16\pi^2)\gamma^{(1)}\ \mathbf{\bar{\Phi}}^L \mathbf{\Phi}_L\ ,
\end{equation}
which vanishes in the conformal limit.
This implies that there is \textit{no} UV divergence in
the integrals appearing in Eqn. (\ref{Keff}). The appearance of the
one-loop $\gamma$ function in the $n=1$ term is also not surprising
since this is the term associated with the one-loop wavefunction
renormalization. This will be true at higher orders as well.
The trihedral symmetry also predicts that the quadratic correction to
the K\"ahler potential will
always be proportional $\mathbf{\bar{\Phi}}^L \mathbf{\Phi}_L$ due to
the diagonal nature of the wavefunction renormalization.

The terms with $n>1$ in Eqn. (\ref{Keff}) are UV finite but are IR
divergent. These are clearly suppressed by suitable powers of the UV
cutoff and disappear in the conformal limit. The trihedral symmetry
also imposes (less stringent)
restrictions on the terms that can appear in these terms. We do not
pursue this here.

\section{Concluding Remarks}

In this paper, we have shown that the trihedral group continues 
to remain a symmetry to two-loops in the quantum theory. We 
conjecture that it is a true symmetry of the LS deformed 
$\mathcal{N}=4$ SYM theory. We also find a interesting 
relationship between holomorphy at two-loop and conformal 
invariance at three-loop -- this appears due to the similarity in 
the coupling constant redefinitions. Ideally, one would like to 
think that the two are indeed the same. But the mismatch of a 
factor of two that we obtain seems to indicate a potential 
conflict. This mismatch can go away in two different ways -- the 
three-loop anomalous dimension computation may be off by a factor 
of two or the two-loop superpotential may be incorrect. Given 
that the diagrams in question do not involve any gauge fields, 
these issues can be addressed in the context of Wess-Zumino 
model. We carried out a detailed investigation of the literature 
in this context and interestingly discovered, in the context of 
the anomalous dimension, two different sets of results. Our 
conclusion is that the results of Jack, Jones and North (derived 
from the result of Parkes) is indeed correct. This leaves open 
the possibility that there is may be a factor of two error in our 
two-loop superpotential. We have been unable to find such an 
error and thus leave this issue for the future.

\noindent \textbf{Acknowledgements} We thank the anonymous referee 
of an earlier paper whose remarks lead us to the investigate the 
quantum aspects of the trihedral symmetry. We would also like to 
thank Justing David and in particular,
Profs. Ian Jack and Tim Jones for extensive email 
correspondence regarding their work. 
The work of KM is
supported by a Senior Research Fellowship from the CSIR (Award No.
9/84(327)/2001-EMR-I).

\appendix

\section{Notations and conventions}

We follow the notation of \cite{Argurio:2003ym} through out this 
paper. The Greek indices $\mu,\nu \ldots = 0,1,2,3$ denote the 
space-time components and $\alpha,\beta = 1,2$ and 
$\dot{\alpha},\dot{\beta} = 1,2$ are the $SU(2)$ spinor indices. 
The $i,j,k,\ldots = 1,2,3$ run over the $SU(3)$ flavor indices 
and $a,b,c, \ldots = 1,...,(N^2-1)$ are the $SU(N)$ color 
indices.  The indices $I,J,K,\ldots$ is a combined notation for 
the flavor and color combination (i,a). The Minkowski metric is 
$g_{\mu \nu} = \mathrm{diag}(+,-,-,-)$. Through out, we use the 
Weyl representation for the spinors. The undotted and dotted 
indices represent chiral and anti-chiral spinors. Spinors are 
raised or lowered as $\psi^\alpha = \epsilon^{\alpha \beta} 
\psi_\beta$, $\psi_\alpha = \epsilon_{\alpha \beta} \psi^\beta$, 
$\psi^{\dot{\alpha}} = \epsilon^{\dot{\alpha} \dot{\beta}} 
\psi_{\dot{\beta}}$, $\psi_{\dot{\alpha}} = 
\epsilon_{\dot{\alpha} \dot{\beta}} \psi^{\dot{\beta}}$, $\alpha 
= 1,2$. Here $\epsilon_{\alpha \beta}$, $\epsilon_{\dot{\alpha} 
\dot{\beta}}$ are totally anti-symmetric tensors. The spinor 
summation convention is
\begin{equation}
\psi \chi = \psi^\alpha  \chi_\alpha \, ; \qquad  
\bar{\psi} \bar{\chi} = \bar{\psi}^{\dot{\alpha}}  \bar{\chi}_{\dot{\alpha}}
\end{equation}
The square of a spinor is 
\begin{equation}
\psi^2 = \frac{1}{2} \psi^\alpha \psi_\alpha \, ; \qquad 
\bar{\psi}^2 = \frac{1}{2} \bar{\psi}_{\dot{\alpha}}  \bar{\psi}^{\dot{\alpha}}
\end{equation}
The derivative with respect to the Grassmann coordinate is defined as
\begin{equation}
\partial_\alpha = \frac{\partial}{\partial \theta^\alpha} \, ; \qquad
\partial_{\dot{\alpha}} = \frac{\partial}{\partial \bar{\theta}^{\dot{\alpha}}}
\end{equation}
The sigma matrices are
\begin{eqnarray}
\sigma_0 = \bar{\sigma}^0 = \left( \begin{array}{c c}-1 & 0 \\
0 & -1
\end{array} \right),\quad 
\sigma_1 = - \bar{\sigma}^1 = \left( \begin{array}{c c}0 & 1 \\
1 & 0
\end{array} \right) \nonumber \\
\sigma_2 = -\bar{\sigma}^2 = \left( \begin{array}{c c}0 & -i \\
i & 0
\end{array} \right),\quad 
\sigma_3 = - \bar{\sigma}^3 = \left( \begin{array}{c c}1 & 0 \\
0 & -1
\end{array} \right) 
\end{eqnarray}
The superspace derivatives are
\begin{equation}
D_\alpha = \partial_\alpha + \frac{i}{2} \sigma^\mu_{\alpha \dot{\alpha}} 
\bar{\theta}^{\dot{\alpha}} \partial_\mu \, ; \qquad 
\bar{D}_{\dot{\alpha}} = \bar{\partial}_{\dot{\alpha}} + \frac{i}{2} 
\theta^{\alpha} \sigma^\mu_{\alpha \dot{\alpha}} \partial_\mu
\end{equation}
obeying the anti-commutation relation
\begin{equation}
\{D_\alpha, \bar{D}_{\dot{\alpha}} \} = i \sigma^\mu_{\alpha \dot{\alpha}} \partial_\mu .
\end{equation}
Further, $D^2 = - \frac12 D^\alpha D_\alpha$ and 
$D^2 = - \frac12 \bar{D}_{\dot{\alpha}} \bar{D}^{\dot{\alpha}}$.

\noindent
The integral over the Grassmann coordinates are defined such that
\begin{equation}
\int d^2 \theta \ \theta^2 = \int d\theta^2 d\theta^1 \; \theta^1 \theta^2
= 1 = \int d^2 \bar{\theta} \ \bar{\theta}^2
\end{equation}

\begin{eqnarray}
\bar{D}_1^2 D_1^2 [q] \delta^4(\theta_{12}) \Big|_{\theta_1=\theta_2} =
1  \\
D_1^2 \bar{D}_1^2 \bar{D}_1^2 D_1^2 [q] \delta^4(\theta_{12})
\Big|_{\theta_1=\theta_2} =  q^2 
\end{eqnarray}

\section{Trace formulae for SU(N)}
Below, we provide the trace identities and normalisations that we have
used in our paper.
\begin{align}\label{fgids}
&T^a T^a = \frac{N^2 - 1}{N} I \qquad  \qquad \textrm{Tr}(T^a T^b) 
= \delta^{ab} \nonumber \\
&\textrm{Tr}(A T^a B T^a) = \textrm{Tr}(A)  \textrm{Tr}(B) 
- \frac1N \textrm{Tr}(A B)  \\
&\textrm{Tr}(A T^a) \textrm{Tr}(B T^a) = \textrm{
Tr}(A B) 
- \frac1N \textrm{Tr}(A) \textrm{Tr}(B)  \nonumber
\end{align}
The following identities are useful in computing the one-loop anomalous
dimension.
\begin{align}
\epsilon^{ikl}\epsilon_{jkl} &= 2 \delta^i_j \\
\bar{c}^{ikl} c_{jkl}&=\big(2|c_0|^2+|c_1|^2 \big)\delta^i_j\\
f^{acd}f_{bcd}&=2N\delta^a_b  \\
d^{acd}d_{bcd}&= \left(\tfrac{N^2-4}{2N}\right) \delta^a_b
\end{align}
The following identities involving five $d$/$f$ tensors are
required in the evaluation of $\mathcal{M}^{IJK}$:
\begin{align}
\label{fiveids}
d^{a_1a_2a_3}d^{b_1b_2b_3}d_{c_1a_1b_1}d_{c_2a_2b_2}d_{c_3a_3b_3}
&=-\frac{N^2-10}{N^2}\ d_{c_1c_2c_3} \nonumber \\
i\ f^{a_1a_2a_3}d^{b_1b_2b_3}d_{c_1a_1b_1}d_{c_2a_2b_2}d_{c_3a_3b_3}
&=-\frac{N^2-4}{2N^2}\ i\ f_{c_1c_2c_3}  \\
(i)^2\ f^{a_1a_2a_3}f^{b_1b_2b_3}d_{c_1a_1b_1}d_{c_2a_2b_2}d_{c_3a_3b_3}
&=2\ d_{c_1c_2c_3}  \nonumber \\
(i)^2\ d^{a_1a_2a_3}d^{b_1b_2b_3}f_{c_1a_1b_1}f_{c_2a_2b_2}d_{c_3a_3b_3}
&=2\ d_{c_1c_2c_3}\nonumber
\end{align}
All other combinations involving five $d$/$f$ tensors 
are \textit{vanishing}. 

\section*{Deriving the identities}

We now sketch the method that we used to derive the various identities
given in Eqn. (\ref{fiveids}. 
In the following, we represent $\textrm{Tr}(T_aT_bT_c)$ by 
$(abc)$. Further, we define
\begin{equation}
(\overline{abc})=\frac12\Big[(abc)+(acb) \Big] \quad,\quad 
(\widetilde{abc})=\frac12\Big[(abc)-(acb) \Big] \ .
\end{equation} 
Thus one has $d_{abc}=(\overline{abc})$ and 
$f_{abc}=\tfrac{2}{i}(\widetilde{abc})$.
Let 
\begin{equation*}
 [00000]_{klm}\equiv (a_1a_2a_3)(b_1b_2b_3)(ka_1b_1)(la_2b_2)(ma_3b_3). 
\end{equation*}

We represent the $32=2^5$ combinations that can appear by a five 
bit number $[c_1c_2c_3c_4c_5]$ with the above equation defining 
$[00000]$. Each of the bits represents the five terms that 
appears in the RHS of the above equation. For instance, $c_1=0$ 
represents $(a_1a_2a_3)$ and $c_1=1$ represents $(a_1a_3a_2)$ and 
so on. There are symmetries which enables us to reduce the 
computation to only four independent terms which we then compute. 
The symmetries are as follows
\begin{enumerate}
\item 
$[c_1c_2c_3c_4c_5]_{klm}=[c_1c_2c_5c_3c_4]_{mkl}=[c_1c_2c_4c_5c_3]_{lmk}$.
\item 
$[c_1c_2c_3c_4c_5]_{klm}=[c_2c_1c_3\oplus1c_4\oplus1c_5\oplus1]_{klm}$ 
where $c_1\oplus 1$ refers to the Boolean operation 
\textit{exor}.
\item $[c_1c_2c_3c_4c_5]_{klm}=[c_1\oplus1c_2\oplus1c_3c_4c_5]_{kml}$.
\end{enumerate}
Further isotropy of $[c_1c_2c_3c_4c_5]_{klm}$ under $SU(N)$ gauge 
transformations implies that
$$
[c_1c_2c_3c_4c_5]_{klm}=A[c_1c_2c_3c_4c_5]\ 
(\overline{klm})+B[c_1c_2c_3c_4c_5]\ (\widetilde{klm})\ ,
$$
where $A[c_1c_2c_3c_4c_5]$ and $B[c_1c_2c_3c_4c_5]$ are 
constants. The symmetries imply that we need to work out only 
four terms: $[00000]$, $[10001]$, $[10000]$, and $[10001]$.
Using the identities given in Eqn. (\ref{fgids}), we obtain
\begin{align*}
[00000]_{klm}&=\Big[1 +\frac{10}{N^2}\Big]\ (\overline{klm})+
\Big[-1+\frac4{N^2}\Big]\ (\widetilde{klm}) \\
[00001]_{klm} &=\Big[-1 +\frac{10}{N^2}\Big]\ (\overline{klm})+
 \Big[-1+\frac4{N^2}\Big]\ (\widetilde{klm}) \\
[10000]_{klm}&=\frac{10}{N^2}\ (\overline{klm}) \\
[10001]_{klm}&=\Big[-2+\frac{10}{N^2}\Big]\ (\overline{klm})
\end{align*}
Using the above four relations we can work out all the 32 
combinations. We can derive identities involving five combinations of
the $d$ and $f$ $SU(N)$ tensors with this information. For instance,
one has in order to obtain the identity involving five $d$ tensors, we
need to compute
$$
\frac1{32}\sum_{c_1,\ldots,c_5} A[c_1c_2c_3c_4c_5]\quad\textrm{and}\quad
\frac1{32}\sum_{c_1,\ldots,c_5} B[c_1c_2c_3c_4c_5]\ .
$$
This is easily done using symbolic manipulation programs such a
Maple/Mathematica. 

\section{Evaluation of integrals}

Here we provide the
details of the computation of Feynman diagrams in Figure \ref{EAD} and Figure \ref{EAF}.
Figure \ref{EAD}a gives the following integral.
\begin{align}
& \int d^2\theta_1 d^4\theta_2 d^4\theta_3 \int \frac{d^Dq}{(2 \pi)^{D}}
\frac{\Phi_I(-p_2 - p_3, \theta_1) \Phi_J(p_2 , \theta_2) \Phi_K(p_3, \theta_3)}{q^2 (q-p_2)^2
(q-p_2-p_3)^2} \nonumber \\
&\left(- 1\right)\; \bar{D}_1^2 D_2^2 [q] \delta^4(\theta_{12})\; 
 \bar{D}_1^2 D_3^2 [q+p_1] \delta^4(\theta_{13}) \; \delta^4(\theta_{32}) \ .
\end{align} 
We convert all the Grassmann integrations over $d^2\theta$ and $d^2\bar{\theta}$ 
into $d^4\theta$ 
by using up factors of $\bar{D}^2$ and $D^2$ respectively and integrate the $\delta$-functions
out. 
\begin{align}
&  \int d^4\theta_1 d^4\theta_2  \int \frac{d^Dq}{(2 \pi)^{D}}
\frac{ \Phi_I(-p_2 - p_3, \theta_1)\ D_2^2 \left( \Phi_J(p_2 , \theta_2) 
\Phi_K(p_3, \theta_2) \right)}{q^2 (q-p_2)^2
(q-p_2-p_3)^2} \nonumber \\
&\left(- 1\right)\; \delta^4(\theta_{12})\; 
 \bar{D}_1^2 D_2^2 [q+p_1] \delta^4(\theta_{12}) \nonumber \\
&= - \int d^2\theta  \int \frac{d^Dq}{(2 \pi)^{D}}
\frac{ \Phi_I(-p_2 - p_3, \theta)\ \bar{D}^2 D^2 \left( \Phi_J(p_2 , \theta) 
\Phi_K(p_3, \theta) \right)}{q^2 (q-p_2)^2
(q-p_2-p_3)^2} \nonumber \\
&= - \int d^2\theta \ \Phi_I(-p_2 - p_3, \theta) \Phi_J(p_2 , \theta) 
\Phi_K(p_3, \theta)   \nonumber \\
& \qquad \qquad \qquad \qquad  \int \frac{d^Dq}{(2 \pi)^D} \frac{p_1^2}{q^2 (q-p_2)^2
(q-p_2-p_3)^2} 
\end{align}
We have simplified the expressions involving $D$-operator,
using the identities given in appendix A.
Figure \ref{EAD}b contributes the integral
\begin{align}
& \int d^2\theta_1 d^4\theta_2 d^4\theta_3 d^4\theta_4 d^4\theta_5 
\int \frac{d^Dk d^Dq}{(2 \pi)^{2D}}
\frac{\Phi_I(-p_2 - p_3, \theta_1) \Phi_J(p_2 , \theta_2) 
\Phi_K(p_3, \theta_3)}{k^2 q^2 (k-q)^2 (q-p_2)^2 (q+p_1)^2 (k+p_1)^2} \nonumber \\
& \bar{D}_1^2 D_4^2 [k] \delta^4(\theta_{14})\; 
 \bar{D}_4^2 D_2^2 [q] \delta^4(\theta_{24}) \; \bar{D}_5^2 D_3^2 [q+p_1] \; \delta^4(\theta_{35})
\; \delta^4(\theta_{23}) \; \delta^4(\theta_{45}) \nonumber \\
&\qquad \qquad \qquad \qquad \qquad \qquad \bar{D}_1^2 D_5^2 [k+p_1] \delta^4(\theta_{15})  
\end{align} 
which is again simplified as above to obtain
\begin{align}
& \int d^2\theta \ \Phi_I(-p_2 - p_3, \theta) \Phi_J(p_2 , \theta) 
\Phi_K(p_3, \theta)   \nonumber \\
& \int \frac{d^Dq}{(2 \pi)^D} \left[ \frac{p_1^4}{k^2 
q^2 (k-q)^2 (q-p_2)^2 (q+p_1)^2 (k+p_1)^2} - \frac{p_1^2}{k^2 
q^2 (k-q)^2 (q-p_2)^2  (k-p_2)^2}\right]
\end{align} 
The blob in Figure \ref{EAD}c consists of a pure chiral superfield loop as
well as one involving gluons. The diagram with a blob (loop) made up of two gluons
and one chiral propagators contributes
\begin{align}
& \int d^2\theta_1 d^4\theta_2 d^4\theta_3 d^4\theta_4 d^4\theta_5 
\int \frac{d^Dk d^Dq}{(2 \pi)^{2D}}
\frac{\Phi_I(-p_2 - p_3, \theta_1) \Phi_J(p_2 , \theta_2) 
\Phi_K(p_3, \theta_3)}{k^2 q^2 (k-q)^2 (q-p_2)^2 (k-p_2)^2 (k+p_1)^2} \nonumber \\
&\left(-1\right) \bar{D}_1^2 D_4^2 [k] \delta^4(\theta_{14})\; 
 \bar{D}_1^2 D_3^2 [k+p_1] \delta^4(\theta_{13}) \; \bar{D}_4^2 D_2^2 [q] \; \delta^4(\theta_{24})
\; \delta^4(\theta_{25}) \; \delta^4(\theta_{35}) \; \delta^4(\theta_{45}) \nonumber \\
\end{align} 
which easily reduces to
\begin{align}
& -p_1^2 \int d^2\theta \
 \int \frac{d^Dq}{(2 \pi)^D}   \frac{\left(\Phi_I(-p_2 - p_3, \theta) \Phi_J(p_2 , \theta) 
\Phi_K(p_3, \theta)  \right)}{k^2 
q^2 (k-q)^2 (q-p_2)^2  (k-p_2)^2 (k+p_1)^2}
\end{align} 
Contribution from Figure \ref{EAD}d is the integral
\begin{align}
& \int d^2\theta_1 d^4\theta_2 d^4\theta_3 d^4\theta_4  
\int \frac{d^Dk d^Dq}{(2 \pi)^{2D}}
\frac{\Phi_I(-p_2 - p_3, \theta_1) \Phi_J(p_2 , \theta_2) 
\Phi_K(p_3, \theta_3)}{k^2 q^2 (k-q-p_2)^2 (q-p_3)^2  (k+p_1)^2} \nonumber \\
& \bar{D}_1^2 D_2^2 [k] \delta^4(\theta_{12})\; 
 \bar{D}_1^2 D_4^2 [k+p_1] \delta^4(\theta_{14}) \; \bar{D}_4^2 D_3^2 [q-p_3] \; 
 \delta^4(\theta_{34})
\; \delta^4(\theta_{24}) \; \delta^4(\theta_{23}) \nonumber \\
\end{align} 
which when simplified reduces to
\begin{align}
& \int d^2\theta \ \Phi_I(-p_2 - p_3, \theta) \Phi_J(p_2 , \theta) 
\Phi_K(p_3, \theta)  
 \int \frac{d^Dq}{(2 \pi)^D}   \frac{p_3^2}{k^2 
q^2 (k-q)^2 (q-p_3)^2  (k-p_3)^2}
\end{align}

The details of the evaluation of the integral from Figure \ref{EAF}
is given below as the exact value of this integral is very crucial.
\begin{align}
&\int d^4\theta_1 d^4\theta_2 d^4\theta_3 \int \frac{d^Dk d^Dq}{(2 \pi)^{2D}}
\frac{\Phi_I(p_2 + p_3, \theta_1) \Phi_J(-p_2 , \theta_2) \Phi_K(-p_3, \theta_3)}{k^2 
q^2 (k-q)^2 (q-p_2)^2
(k-q-p_2)^2 (k-p_2-p_3)^2} \nonumber \\
& \bar{D}_2^2 D_1^2 [q] \delta^4(\theta_{12})\; 
 D_1^2 [k-q] \delta^4(\theta_{13}) 
\; \bar{D}_1^2 D_2^2 [k-p_1] \delta^4(\theta_{12}) \; \bar{D}_3^2 D_2^2 [q] \delta^4(\theta_{32}) \ .
\end{align} 
We can integrate the $D$-operators by parts and simplify this by getting rid of 
the Grassmann integrals one by one.  
\begin{align}
&\int d^4\theta_1 d^4\theta_2  \int \frac{d^Dk d^Dq}{(2 \pi)^{2D}}
\frac{D_1^2 \Phi_I(p_2 + p_3, \theta_1) \Phi_J(-p_2 , \theta_2) \Phi_K(-p_3, \theta_1)}{k^2 q^2 
(k-q)^2 (q-p_2)^2 (k-q-p_2)^2 (k-p_2-p_3)^2} \nonumber \\
& D_2^2 \bar{D}_2^2 \bar{D}_1^2 D_1^2 [q] \delta^4(\theta_{12})\;   
\delta^4(\theta_{12})\; \bar{D}_1^2 D_2^2 [k-p_1] \delta^4(\theta_{12}) 
\end{align}
Using the identities in appendix A
and rewriting the integral over $d^4\theta$ as a chiral integral
\begin{equation}
\int d^2\theta  \int \frac{d^Dk d^Dq}{(2 \pi)^{2D}}
\frac{\bar{D}^2 D^2 \Phi_I(p_2 + p_3, \theta) \; \Phi_J(-p_2 , \theta) \; 
\Phi_K(-p_3, \theta)}{k^2 q^2 (k-q)^2 (q-p_2)^2 (k-q-p_2)^2 (k-p_2-p_3)^2} \nonumber
\end{equation}
Setting $p_2 = 0$ and re-labelling $r = k-q$ and using 
$\bar{D}^2 D^2(p) \Phi(p,\theta) = p^2 \Phi(p,\theta)$,
\begin{eqnarray}
&& \int d^2\theta\;  p_3^2 \; 
\int \frac{d^Dk d^Dq}{(2 \pi)^{2D}}
 \frac{\Phi_I(p_3, \theta) \Phi_J(p_3 , \theta) \Phi_K(-p_3, \theta)
}{k^2 r^2  (k-r)^2 (r-p_3)^2 (k-p_3)^2} \nonumber \\
&=&\mathcal{K}\ \int d^2\theta   \; \Phi_I(p_3, \theta) \Phi_J(p_3 , \theta) \Phi_K(-p_3, \theta)
%=\ \frac{1}{6} \frac{3 M^{IJK} \zeta(3)}{(16 \pi^2)^2}  \int d^2\theta \; \Phi_I \Phi_J \Phi_K
\end{eqnarray}
where $\mathcal{K}$ is the finite integral
\begin{equation}
\mathcal{K}\equiv 
p_3^2\int \frac{d^Dk d^Dq}{(2 \pi)^{2D}}
\frac1{k^2 r^2  (k-r)^2 (r-p_3)^2 (k-p_3)^2} 
= \frac{\kappa}{(16 \pi^2)^2}\ 
\end{equation}
with $\kappa=6\zeta(3)$.

\end{document}